# Identity of King and Flajolet & al. Formulae for LRU Miss Rate Exact Computation

**Christian BERTHET**
STMicroelectronics, Grenoble, France,

**Abstract**. This short paper gives a detailed proof of identity between two classic formulas for the computation of the exact Miss Rate of LRU caches. An extension to the identity of two formulas of the expected time of a partial collection in the coupon collector problem is also presented.

## Introduction

There exist two formulae for the exact computation of LRU Miss Rate. As early as 1971 W.F. King derived a formula to compute the fault probability ('miss rate') of LRU demand paging systems [King71]. And, in 1992, Flajolet et al. gave an integral formula of that probability [Flajolet92]. In their paper, Flajolet et al. mention that "King gave another form of that probability", implying that they used a different method for counting the same thing. However, to our knowledge, a formal equivalence between these two formulae has never been established yet. In this paper we show a *direct* equivalence between the two by algebraic means.

In both cases, hypotheses on the LRU system are the following: system is assumed to obey an Independent-Reference Model (IRM), i.e. cache accesses are independent of past history. Addresses of cache lines (or 'pages') are characterized by a 'popularity' distribution, i.e. a general probability law (not necessarily uniform). Finally, for readers familiar with HW design, the cache is fully associative (in other words, dealing with limited associativity such as in HW caches require the need of resorting to other models).

In this short paper, we show the equivalence of the two afore-mentioned LRU Miss Rate formulas. Proof is based on a simple rewriting mechanism that makes use of a combinatorial identity on the permutations of a subset of probabilities proven in Lemma 1.
In a final section, we give an extension to the equivalence of two other, and similar, formulas for the exact computation of the expected time of a partial collection in the coupon collector problem under a general probability law.

# King Formula for LRU MR

*Notation*

We assume a probability law with general (non-uniform) distribution $p_i, i \in \{1..m\}, \sum_{i=1}^{m} p_i = 1$.
Probability law is often called 'popularity' of the addresses $\{1,..,m\}$.

*Original Formula*

King formula for the Miss Rate of an LRU cache of **capacity** j, $1<j<m$ is [King71] as follows

$$King[j] = \sum_{\substack{j-tuple\{i_1,i_2,...i_j\} \\ i_1 \neq i_2 \neq .. \neq i_j}} \frac{p_{i_1} p_{i_2} .. p_{i_{j-1}} p_{i_j} (1 - p_{i_1} - p_{i_2} .. - p_{i_j})}{(1-p_{i_1})(1-p_{i_1}-p_{i_2})..(1-p_{i_1}-p_{i_2}..-p_{i_{j-1}})}$$

Miss rate is a summation over all the **t-uples** of size j, i.e. all the **permutations** of all the **subsets** of size j of the set $\{1, .., m\}$. This Formula is also given in Fagin papers [Fagin77, Fagin78].

*Rewriting of King Formula*

Let J be a subset of the set $\{1,...,m\}$, with $j=|J|<m$, we introduce the notation:

$$I_J = \sum_{\substack{permutation \\ \{i_1,i_2,...i_j\} of J}} \frac{p_{i_1} p_{i_2} .. p_{i_j}}{(1-p_{i_1})(1-p_{i_1}-p_{i_2})..(1-p_{i_1}-p_{i_2}..-p_{i_j})}$$

Noticing that $\sum_{i \in J} p_i$ is independent of the permutation of the subset J, King initial formula is rewritten: $King[j] = \sum_{|J|=j} \left( \left(1 - \sum_{i \in J} p_i\right)^2 I_J \right) = \sum_{|J|=j} \left( \left(\sum_{i \notin J} p_i\right)^2 I_J \right)$.

**Flajolet Integral Formula**

*Original formula*

An integral formula of the LRU Miss rate of a cache of size j is given by Flajolet et al. ([Flojolet92] Theorem 5.1 pg. 219):

$$1 - MR[j] = \sum_{q=0}^{j-1} [u^q] \int_0^\infty \left( \prod_{i=1}^{m} (1 + u(e^{p_i t} - 1)) \right) \left( \sum_{i=1}^{m} \frac{p_i^2}{1 + u(e^{p_i t} - 1)} \right) e^{-t} dt$$

where $[u^j]f(u)$ denotes the coefficient of $u^j$ in the polynomial $f(u)$. It can be re-written using successively variable change $x=e^{-t}$, distribution of the sum, computation of $u^j$ coefficient and distribution of the integral:

$$1-MR[j] = \sum_{q=0}^{j-1}[u^q]\int_0^1\left(\prod_{i=1}^m(1+u(x^{-p_i}-1))\right)\left(\sum_{i=1}^m\frac{p_i^2}{1+u(x^{-p_i}-1)}\right)dx = \sum_{q=0}^{j-1}\int_0^1\left(\sum_{i=1}^m[u^q]p_i^2\left(\prod_{\substack{k=1\\k\neq i}}^m(1+u(x^{-p_k}-1))\right)\right)dx$$

$$= \sum_{q=0}^{j-1}\int_0^1\left(\sum_{i=1}^m p_i^2 \sum_{\substack{|K|=q\\i\notin K}}\left(\prod_{k\in K}(x^{-p_k}-1)\right)\right)dx = \sum_{q=0}^{j-1}\sum_{i=1}^m p_i^2 \sum_{\substack{|K|=q\\i\notin K}}\int_0^1\left(\prod_{k\in K}(x^{-p_k}-1)\right)dx$$

We now define the following lemma which is proven in Appendix 1. It allows us to relate the two MR formulae using the quantity $I_J$ defined above.

*Lemma 1*

Let J be a subset J of size j of the set $\{1,\ldots,m\}$, $1\leq j<m$. It holds that: $I_J = \int_0^1\left(\prod_{i\in J}(x^{-p_i}-1)\right)dx$

*Rewriting of Flajolet et al. formula*

Using Lemma 1, Flajolet et al. formula is:

$$1-MR[j] = \sum_{q=0}^{j-1}\sum_{i=1}^m p_i^2 \sum_{\substack{|K|=q\\i\notin K}}I_K = \sum_{i=1}^m p_i^2 \sum_{\substack{0\leq|K|<j\\i\notin K}}I_K = \sum_{0\leq|K|<j}I_K\left(\sum_{i\notin K}p_i^2\right)$$

Using the fact that, for $j=1$, $1-MR[1] = \sum_{i=1}^m p_i^2$ (obviously a cache of capacity 1 hits only for successive accesses to the same address) $I_J$ notation can be extended to null set with: $I_\emptyset = 1$.

Also since Miss Rate is null for a capacity $j=m$, it holds that $\sum_{0\leq|K|<m}I_K\left(\sum_{i\notin K}p_i^2\right)=1$, so MR[j] can also be expressed as follows: $\sum_{j\leq|J|<m}I_J\left(\sum_{i\notin J}p_i^2\right)$.

At first sight, King formula may seem more tractable than that of Flajolet et al. since the summation is performed on a single value j and not on a subset [0,j[ or [j,m[. However this is very arguable since both expressions require the enumeration of all subsets which is intractable.

**Equivalence of King and Flajolet formulas**

Equivalence of these two formulas reduces to proving that: $\sum_{|J|=j}\left(\sum_{i\notin J} p_i\right)^2 I_J = 1 - \sum_{0\leq |J|<j} I_J \left(\sum_{i\notin J} p_i^2\right)$

*Uniform Distribution*

Let us first reassure ourselves all this makes sense with a simple uniform distribution ($p_i=1/m$ for all i). Using variable change $X=x^{1/m}$ (i.e., $mX^{m-1}dX=dx$), it follows that:

$$I_J = \int_0^1 (x^{-1/m} - 1)^j dx = \int_0^1 (X^{-1} - 1)^j m X^{m-1} dX = m\int_0^1 (1-X)^j X^{m-1-j} dX = m\frac{j!(m-j-1)!}{m!} = \binom{m-1}{j}^{-1}.$$

Relation $\int_0^1 x^a (1-x)^b dx = \frac{a!b!}{(a+b+1)!}$ is classically obtained using iterated integration by parts.

King Formula collapses to the well-known LRU MR value for a uniform distribution:

$$\sum_{|J|=j}\left(\sum_{i\notin J} p_i\right)^2 I_J = \binom{m}{j}\left(1-\frac{j}{m}\right)^2 \binom{m-1}{j}^{-1} = 1-\frac{j}{m}.$$

And Flajolet et al. on the other hand:

$$1 - MR[j] = \sum_{0\leq |J|<j} I_J\left(\sum_{i\notin J} p_i^2\right) = \sum_{q=0}^{j-1}\sum_{|J|=q} I_J\left(\sum_{i\notin J} p_i^2\right) = \sum_{q=0}^{j-1}\binom{m}{q}\binom{m-1}{q}^{-1}\frac{(m-q)}{m^2} = \frac{1}{m}\sum_{q=0}^{j-1} 1 = \frac{j}{m}.$$

Before proving the equivalence in the general case, we first introduce the following lemma which is proven in Appendix 2.

*Lemma 2*

Let J be a subset of the set {1,…,m} of size j<m, and F be a numerical function. It holds that:
$$\sum_{|J|=j}\sum_{i\notin J} F(i,J) = \sum_{|J|=j+1}\sum_{i\in J} F(i, J-\{i\})$$

*Equivalence of King and Flajolet formulas in general case*

We give the proof of equivalence for a general distribution: $\sum_{|J|=j}\left(\sum_{i\notin J} p_i\right)^2 I_J = 1 - \sum_{0\leq |J|<j} I_J\left(\sum_{i\notin J} p_i^2\right)$

Identity is obvious for |J|=1 since: $\sum_{j=1}^n (1-p_j)^2 \frac{p_j}{1-p_j} = \sum_{i=1}^m p_i - \sum_{i=1}^m p_i^2 = 1 - I_\emptyset \sum_{i=1}^m p_i^2$.

Also note that for j=m-1: $\sum_{|J|=m-1}(p_{i\notin J})^2 I_J = \sum_{m-1\leq|J|<m} I_J(p_{i\notin J}^2)$.

For the general case we proceed by induction on the size j:

$$MR[j+1] = 1 - \sum_{0\leq|J|<j+1} I_J \sum_{i\notin J} p_i^2 = MR[j] - \sum_{|J|=j} I_J \sum_{i\notin J} p_i^2$$

From the hypothesis MR[j]=King[j], and noting $q_J = \sum_{i\notin J} p_i$, induction step is:

$$MR[j+1] = \sum_{|J|=j}\left(\left(\sum_{i\notin J} p_i\right)^2 - \sum_{i\notin J} p_i^2\right) I_J = \sum_{|J|=j}\left(\sum_{i\notin J} p_i \left(\sum_{\substack{k\notin J \\ k\neq i}} p_k\right)\right) I_J = \sum_{|J|=j}\left(\sum_{i\notin J} p_i(q_J - p_i)\right) I_J$$

On the other hand: $King[j+1] = \sum_{|J|=j+1}\left(\sum_{i\notin J} p_i\right)^2 I_J = \sum_{|J|=j+1} q_J^2 I_J$

From the definition of $I_J$ the following relation holds: $q_J I_J = \sum_{i\in J} p_i I_{J-\{i\}}$. This stems directly from considering the permutations of J whose last element corresponds to the sum index 'i'.

Consequently $King[j+1] = \sum_{|J|=j+1}\sum_{i\in J} q_J p_i I_{J-\{i\}} = \sum_{|J|=j+1}\sum_{i\in J}(q_{J-\{i\}} - p_i)p_i I_{J-\{i\}}$

Lemma 2 applied to the function $F(i, J-\{i\}) = (q_{J-\{i\}} - p_i)p_i I_{J-\{i\}}$ finally leads to the desired result: $King[j+1] = \sum_{|J|=j}\sum_{i\notin J}(q_J - p_i)p_i I_J = MR[j+1]$. QED.

## Application to CCP expectation of a partial collection

*Flajolet et al. Integral Formula for CCP expectation*

For a general (non-uniform) distribution $p_i, i \in \{1..m\}, \sum_{i=1}^{m} p_i = 1$, Flajolet et al. have given a formula for the expected waiting time to collect j items out of m ('partial collection') of the coupon-collector problem:

Expected time is ([Flajolet92] 13a p.216): $E\{C_j\} = \sum_{q=0}^{j-1} \int_0^\infty [u^q] \left( \prod_{i=1}^{m} (1 + u(e^{p_i t} - 1)) \right) e^{-t} dt$, where j≤m is the size of the collection and $[u^q]f(u)$ denotes the coefficient of $u^q$ in the polynomial f(u). It is obvious that $E\{C_0\} = 0$ and $E\{C_1\} = 1$.

Let us notice that formula has a simple recurrence relation:

$$E\{C_{j+1}\} - E\{C_j\} = \int_0^\infty [u^j] \left( \prod_{i=1}^{m} (1 + u(e^{p_i t} - 1)) \right) e^{-t} dt \text{ , } 1 \leq j < m,$$

which can be rewritten with variable change $x=e^{-t}$, computation of $u^j$ coefficient and distribution of the integral: $E\{C_{j+1}\} - E\{C_j\} = \sum_{|J|=j} \int_0^1 \left( \prod_{i \in J} (x^{-p_i} - 1) \right) dx$.

In the sequel, we denote this expression $\Delta E\{j\}$. Using Lemma 1 and $I_J$ notation, it follows that:

$\Delta E\{j\} = \sum_{|J|=j} I_J$ and $E\{C_j\} = \sum_{0 \leq |J| < j} I_J$.

Let us notice that, for a uniform distribution: $\Delta E\{j\} = \binom{m}{j} \binom{m-1}{j}^{-1} = \frac{m}{m-j}$ giving the well-known

$E\{C_j\} = m(H_m - H_{m-j})$ where $H_n$ is the n-th harmonic number.

In their paper, Flajolet et al. also give a symmetric functions expression (or a variant of it, after index change)

$$E\{C_j\} = \sum_{k=m-j+1}^{m} (-1)^{j+k-m-1} \binom{k-1}{m-j} \sum_{|J|=k} \frac{1}{P_J}$$

where $P_J$ is the sum of the probabilities of the subset J.

*Alternative formula from Ferrante et al.*

An alternative formula for the expectation of a partial collection, noted $E[X_m(k)]$ – equivalent of Flajolet $E\{C_k\}$ - and based on conditional probabilities is given in Proposition 2 page 7 of paper [Ferrante12]. It is very similar to King Formula for the LRU miss rate calculation since it requires the enumeration of all the permutations of all the subsets. Expectation for a partial collection of size k out of m is defined as: $E[X_m(k)] = \sum_{s=1}^{k} E[X_s]$ where $E[X_s]$ is (using their notation for the permutations):

$$E[X_s] = \sum_{i_1 \neq i_2 \neq .. \neq i_{s-1}=1}^{m} \frac{p_{i_1} p_{i_2} .. p_{i_{s-1}}}{(1-p_{i_1})(1-p_{i_1}-p_{i_2})..(1-p_{i_1}-p_{i_2}..-p_{i_{s-1}})}$$

Again, there is a very simple recurrence relation $E[X_m(k)] = E[X_m(k-1)] + E[X_k]$ (assuming $E[X_1]=1$).

*Equivalence of Flajolet et al. $E\{C_j\}$ and Ferrante&al. $E[X_m(j)]$*

For a uniform probability $E[X_s] = \frac{m!}{(m-s+1)!} \cdot \frac{(m-s)!}{(m-1)!} = \frac{m}{m-s+1}$ , hence

$$E[X_m(k)] = \sum_{s=1}^{k} \frac{m}{m-s+1} = E\{C_k\}.$$

For j=2, whatever m≥2, equivalence is readily obtained by computation of the sums of Flajolet et al. symmetric functions expression:

$$E\{C_2\} = \sum_{k=m-1}^{m} (-1)^{k-m+2-1} \binom{k-1}{m-2} \sum_{|J|=k} \frac{1}{P_J} = \sum_{|J|=m-1} \frac{1}{P_J} + (-1)^1 \binom{m-1}{m-2} \sum_{|J|=m} \frac{1}{P_J} = \sum_{i=1}^{m} \frac{1}{1-p_i} - (m-1)$$

$$= 1 + \sum_{i=1}^{m} \frac{p_i}{1-p_i} = E[X_m(2)]$$

More generally, identity of Flajolet et al. and Ferrante et al. relations for any j≤m can be stated simply by noticing that $E[X_s] = \sum_{|J|=s-1} I_J$ hence $E[X_m(k)] = \sum_{s=1}^{k} \sum_{|J|=s-1} I_J = E\{C_k\}$.  QED.

**Conclusion**

Using algebraic means we have shown the identity of two formulas for the exact computation of LRU miss rate and two other formulas for the expectation of a partial collection in CCP. Both identities make use of a combinatorial relation on the permutations of a subset proved in Appendix 1.

**Acknowledgments**

The author thanks Russell May for his very helpful comments.

**References**

1. W.F. King III, "Analysis of demand paging algorithms", Proc. of the I.F.I.P. Congress 1971, North-Holland Publishing Company.

2. Ronald Fagin, "Asymptotic miss ratios over independent references", Journal of Computer and System Sciences Volume 14, Issue 2, April 1977, Pages 222–250
http://www.sciencedirect.com/science/article/pii/S0022000077800147

3. Fagin, R., & Price, T. G. (1978). Efficient calculation of expected miss ratios in the independent reference model. SIAM Journal on Computing, 7(3), 288-297.

4. Flajolet, Philippe; Gardy, Danièle; Thimonier, Loÿs (1992), "Birthday paradox, coupon collectors, caching algorithms and self-organizing search", Discrete Applied Mathematics 39 (3): 207–229. doi:10.1016/0166-218X(92)90177-C, MR 1189469.

5. Ferrante, M., & Frigo, N. (2012). On the expected number of different records in a random sample. *arXiv preprint arXiv:1209.4592*.

## Appendix 1: Proof of Lemma 1

Let J be a subset J of size j of the set {1,...,m}, 1≤j<m. We want to prove that:

$$\int_0^1 \left(\prod_{i \in J}(x^{-p_i}-1)\right)dx = \sum_{\substack{permutation \\ \{i_1,i_2,...i_j\}\, of\, J}} \frac{p_{i_1} p_{i_2} \cdots p_{i_j}}{(1-p_{i_1})(1-p_{i_1}-p_{i_2})\cdots(1-p_{i_1}-p_{i_2}\cdots-p_{i_j})}$$

For sake of simplicity of notation, we note {1,2,...,j} the subset J.
Identity is easily proven for j=1 and j=2 by expanding and evaluating the integrals.

For |J|=1: $\int_0^1 (x^{-p_1}-1)dx = \frac{p_1}{1-p_1}$ and for |J|=2:

$$\int_0^1 (x^{-p_1}-1)(x^{-p_2}-1)dx = \frac{1}{1-p_1-p_2} - \frac{1}{1-p_1} - \frac{1}{1-p_2} + 1 = \frac{p_1 p_2 (2-p_1-p_2)}{(1-p_1)(1-p_2)(1-p_1-p_2)}$$

$$= \frac{p_1 p_2}{(1-p_1)(1-p_1-p_2)} + \frac{p_2 p_1}{(1-p_2)(1-p_2-p_1)}$$

We prove the general case by induction on j and we note $R_j = \int_0^1 \left(\prod_{i=1}^{j}(x^{-p_i}-1)\right)dx$.

Then: $R_{j+1} = \int_0^1 \left(\prod_{i=1}^{j+1}(x^{-p_i}-1)\right)dx = \int_0^1 \left(\prod_{i=1}^{j}(x^{-p_i}-1)\right)x^{-p_{j+1}}dx - R_j$

We introduce the variable change $z = x^{1-p_{j+1}}$ hence $dz = (1-p_{j+1})x^{-p_{j+1}}dx$ and

$$R_{j+1} = \frac{1}{1-p_{j+1}}\int_0^1 \left(\prod_{i=1}^{j}(z^{-\frac{p_i}{1-p_{j+1}}}-1)\right)dz - \int_0^1 \left(\prod_{i=1}^{j}(x^{-p_i}-1)\right)dx .$$

At this point we assume the induction hypothesis holds for the two integrals, both of them being indexed by j. For the first integral, notice that:

$$\int_0^1 \left(\prod_{i=1}^{j}(z^{-\frac{p_i}{1-p_{j+1}}}-1)\right)dz = \sum_{\substack{permutation \\ \{i_1,i_2,...i_j\}\, of\, J}} \frac{\frac{p_{i_1}}{1-p_{j+1}}\frac{p_{i_2}}{1-p_{j+1}}\cdots\frac{p_{i_j}}{1-p_{j+1}}}{(1-\frac{p_{i_1}}{1-p_{j+1}})(1-\frac{p_{i_1}}{1-p_{j+1}}-\frac{p_{i_2}}{1-p_{j+1}})\cdots(1-\frac{p_{i_1}}{1-p_{j+1}}-\frac{p_{i_2}}{1-p_{j+1}}\cdots-\frac{p_{i_j}}{1-p_{j+1}})}$$

$$= \sum_{\substack{permutation \\ \{i_1,i_2,...i_j\}\, of\, J}} \frac{p_{i_1}p_{i_2}\cdots p_{i_j}}{(1-p_{j+1}-p_{i_1})(1-p_{j+1}-p_{i_1}-p_{i_2})\cdots(1-p_{j+1}-p_{i_1}-p_{i_2}\cdots-p_{i_j})}$$

We use the intermediate variables $\{q_{i_1}=1-p_{i_1}; q_{i_2}=1-p_{i_1}-p_{i_2}; q_{i_j}=1-p_{i_1}-p_{i_2}-\cdots-p_{i_j}\}$ to make the proof more readable, hence: $R_j = \int_0^1 \left(\prod_{i \in J}(x^{-p_i}-1)\right)dx = \sum_{\substack{permutation \\ \{i_1,i_2,...i_j\}\, of\, J}} \frac{p_{i_1} p_{i_2} \cdots p_{i_j}}{q_{i_1} q_{i_2} \cdots q_{i_j}}$

And:

$$R_{j+1} = \left(\frac{1}{1-p_{j+1}}\right) \sum_{\substack{permutation \\ \{i_1,i_2,...i_j\} of J}} \frac{p_{i_1} p_{i_2}..p_{i_j}}{(q_{i_1}-p_{j+1})(q_{i_2}-p_{j+1})..(q_{i_j}-p_{j+1})} - \sum_{\substack{permutation \\ \{i_1,i_2,...i_j\} of J}} \frac{p_{i_1} p_{i_2}..p_{i_j}}{q_{i_1} q_{i_2}..q_{i_j}}$$

$$= \sum_{\substack{permutation \\ \{i_1,i_2,...i_j\} of J}} \left( \frac{p_{i_1} p_{i_2}..p_{i_j}}{(1-p_{j+1})(q_{i_1}-p_{j+1})(q_{i_2}-p_{j+1})..(q_{i_j}-p_{j+1})} - \frac{p_{i_1} p_{i_2}..p_{i_j}}{q_{i_1} q_{i_2}..q_{i_j}} \right)$$

Now, since in a subset of size j+1, there are (j+1) permutations for each permutation of a subset J of size j, we can enumerate all the possible positions of element j+1:

$$\sum_{\substack{permutation \\ \{i_1,i_2,...i_{j+1}\} of \\ J \cup \{j+1\}}} \frac{p_{i_1} p_{i_2}..p_{i_{j+1}}}{q_{i_1} q_{i_2}..q_{i_{j+1}}} = \sum_{\substack{permutation \\ \{i_1,i_2,...i_j\} of J}} \left( \begin{array}{c} \frac{p_{j+1} p_{i_1} p_{i_2}..p_{i_j}}{(1-p_{j+1})(q_{i_1}-p_{j+1})(q_{i_2}-p_{j+1})..(q_{i_j}-p_{j+1})} + \\ \frac{p_{i_1} p_{j+1} p_{i_2}..p_{i_j}}{q_{i_1}(q_{i_1}-p_{j+1})(q_{i_2}-p_{j+1})..(q_{i_j}-p_{j+1})} + \\ \frac{p_{i_1} p_{i_2} p_{j+1}..p_{i_j}}{q_{i_1} q_{i_2}(q_{i_2}-p_{j+1})..(q_{i_j}-p_{j+1})} +..+ \frac{p_{i_1} p_{i_2}..p_{i_j} p_{j+1}}{q_{i_1} q_{i_2}..q_{i_j}(q_{i_j}-p_{j+1})} \end{array} \right)$$

Lemma 1 holds if and only if we can prove that $R_{j+1}$ and $\sum_{\substack{permutation \\ \{i_1,i_2,...i_{j+1}\} of J \cup \{j+1\}}} \frac{p_{i_1} p_{i_2}..p_{i_{j+1}}}{q_{i_1} q_{i_2}..q_{i_{j+1}}}$ are the same

expression, in other words, for each permutation of J, the following holds:

$$\frac{1}{(1-p_{j+1})(q_{i_1}-p_{j+1})(q_{i_2}-p_{j+1})..(q_{i_j}-p_{j+1})} - \frac{1}{q_{i_1} q_{i_2}..q_{i_j}} =$$

$$p_{j+1} \left( \begin{array}{c} \frac{1}{(1-p_{j+1})(q_{i_1}-p_{j+1})(q_{i_2}-p_{j+1})..(q_{i_j}-p_{j+1})} + \\ \frac{1}{q_{i_1}(q_{i_1}-p_{j+1})(q_{i_2}-p_{j+1})..(q_{i_j}-p_{j+1})} + \\ \frac{1}{q_{i_1} q_{i_2}(q_{i_2}-p_{j+1})..(q_{i_j}-p_{j+1})} +..+ \frac{1}{q_{i_1} q_{i_2}..q_{i_j}(q_{i_j}-p_{j+1})} \end{array} \right)$$

Which simplifies to:

$$\frac{1}{(q_{i_1}-p_{j+1})(q_{i_2}-p_{j+1})..(q_{i_j}-p_{j+1})} - \frac{1}{q_{i_1} q_{i_2}..q_{i_j}} =$$

$$p_{j+1} \left( \begin{array}{c} \frac{1}{q_{i_1}(q_{i_1}-p_{j+1})(q_{i_2}-p_{j+1})..(q_{i_j}-p_{j+1})} + \\ \frac{1}{q_{i_1} q_{i_2}(q_{i_2}-p_{j+1})..(q_{i_j}-p_{j+1})} +..+ \frac{1}{q_{i_1} q_{i_2}..q_{i_j}(q_{i_j}-p_{j+1})} \end{array} \right)$$

Hence Lemma 1 holds if and only if:

$$q_{i_1}q_{i_2}..q_{i_j} - (q_{i_1} - p_{j+1})(q_{i_2} - p_{j+1})..(q_{i_j} - p_{j+1}) =$$
$$p_{j+1}\left(q_{i_2}..q_{i_j} + q_{i_3}..q_{i_j}(q_{i_1} - p_{j+1}) + .. + (q_{i_1} - p_{j+1})(q_{i_2} - p_{j+1})..(q_{i_{j-1}} - p_{j+1})\right)$$

Without loss of generality we simply note $\{p_1, p_2, .., p_j\}$ the permutation and consequently $\{q_1, q_2, .., q_j\}$ the intermediate variables.

Noting $X = p_{j+1}$ and with the convention that a product is equal to 1 if it is an empty product (i.e., the index range is empty) previous equality is equivalent to:

$$\prod_{i=1}^{j} q_i - \prod_{i=1}^{j}(q_i - X) = X \sum_{i=1}^{j}\left(\left(\prod_{k=1}^{i-1}(q_k - X)\right)\left(\prod_{k=i+1}^{j} q_k\right)\right).$$

It holds if and only if the polynomial coefficients (noted $[X^n]$ as usual) of both sides are equal $[X^n]RHS = [X^n]LHS$ for any n, $0 \leq n \leq j$.

This is obviously true for n=0 since: $\prod_{i=1}^{j} q_i - \prod_{i=1}^{j} q_i = 0$.

For n=1, we obtain the identity: $-\left(\sum_{i=1}^{j} -\prod_{\substack{k=1 \\ k \neq i}}^{j} q_k\right) = \sum_{i=1}^{j}\left(\left(\prod_{k=1}^{i-1} q_k\right)\left(\prod_{k=i+1}^{j} q_k\right)\right).$

For the general case, one obtains on the left hand side:

$$[X^n]LHS = [X^n](-\prod_{i=1}^{j}(q_i - X)) = (-1)^{n+1} \sum_{\substack{J \subseteq \{1,..j\} \\ |J|=j-n}} \prod_{i \in J} q_i$$

and on the right hand side:

$$RHS = \sum_{i=1}^{j}\left((-(q_i - X) + q_i)\left(\prod_{k=1}^{i-1}(q_k - X)\right)\left(\prod_{k=i+1}^{j} q_k\right)\right)$$
$$= -\sum_{i=1}^{j}\left(\left(\prod_{k=1}^{i}(q_k - X)\right)\left(\prod_{k=i+1}^{j} q_k\right)\right) + \sum_{i=1}^{j}\left(\left(\prod_{k=1}^{i-1}(q_k - X)\right)\left(\prod_{k=i}^{j} q_k\right)\right)$$

The n-th coefficient of the first sum is null for index $i<n$ and that of the second sum is null for $i-1<n$ hence (noting $Q_J = \prod_{i \in J} q_i$ and $Q_i^j = \prod_{k=i}^{j} q_k$):

$$[X^n]RHS = -\sum_{i=n}^{j}(-1)^n\left(\sum_{\substack{J \subseteq \{1,...,i\} \\ |J|=i-n}} Q_J\right) Q_{i+1}^j + \sum_{i=n+1}^{j}(-1)^n\left(\sum_{\substack{J \subseteq \{1,...,i-1\} \\ |J|=i-1-n}} Q_J\right) Q_i^j.$$

Changing the index of the second summation (to i-1) gives:

$$[X^n]RHS = -\sum_{i=n}^{j}(-1)^n\left(\sum_{\substack{J \subseteq \{1,...,i\} \\ |J|=i-n}} Q_J\right) Q_{i+1}^j + \sum_{i=n}^{j-1}(-1)^n\left(\sum_{\substack{J \subseteq \{1,...,i\} \\ |J|=i-n}} Q_J\right) Q_{i+1}^j$$

The two summations differ only by their upper bounds, thus:

$$[X^n]RHS = (-1)^{n+1}\left(\left(\sum_{\substack{J\subseteq\{1,\ldots,j\}\\|J|=j-n}}Q_J\right)Q_{j+1}^j\right) = [X^n]LHS \text{ since by convention } Q_{j+1}^j \text{ is 1.}$$

This completes the proof of identity.                                                                                QED.

**Appendix 2: Proof of Lemma 2**

Let J be a subset J of size j<m of the set {1,...,m}, j<m, and F a numerical function. We want to prove that:
$$\sum_{|J|=j}\sum_{i\notin J}F(i,J) = \sum_{|J|=j+1}\sum_{i\in J}F(i,J-\{i\})$$

For j=1:
$$\sum_{J=\{j\}}\sum_{i\notin J}F(i,J) = \sum_{j=1}^{m}\left(\sum_{\substack{i=1\\j\neq i}}^{m}F(i,\{j\})\right) = \sum_{1\leq i<j\leq m}F(i,\{j\}) + \sum_{1\leq j<i\leq m}F(i,\{j\})$$

$$= \sum_{1\leq i<j\leq m}F(i,\{j\}) + \sum_{1\leq i<j\leq m}F(j,\{i\}) = \sum_{j=1}^{m}\left(\sum_{i=j+1}^{m}F(i,\{j\}) + F(j,\{i\})\right) = \sum_{J=\{i,j\}}\sum_{k\in J}F(k,J-\{k\}).$$

In the general case J={$i_1,i_2,\ldots i_j$}, sum has j+1 terms:

$$\sum_{|J|=j}\sum_{k\notin J}F(k,J) = \sum_{1\leq k<i_1<i_2<\ldots i_j\leq m}F(k,J) + \sum_{1\leq i_1<k<i_2<\ldots i_j\leq m}F(k,J) + \ldots + \sum_{1\leq i_1<i_2<\ldots k<i_j\leq m}F(k,J) + \sum_{1\leq i_1<i_2<\ldots i_j<k\leq m}F(k,J)$$

Then denoting K the set {$i_1,i_2,\ldots i_j,k$}:

$$\sum_{|J|=j}\sum_{k\notin J}F(k,J) = \sum_{|K|=j+1}F(i_1,K-\{i_1\}) + \sum_{|K|=j+1}F(i_2,K-\{i_2\}) + \ldots + \sum_{|K|=j+1}F(i_j,K-\{i_j\}) + \sum_{|K|=j+1}F(k,K-\{k\})$$

$$= \sum_{|K|=j+1}\sum_{i\in K}F(i,K-\{i\}).$$

                                                                                                                         QED.